\documentclass[3p,sort&compress,12pt,review]{elsarticle}

\usepackage[utf8]{inputenc} 
\usepackage{amsmath}
\usepackage{amssymb}
\usepackage{graphicx}
\usepackage{siunitx}
\usepackage{textcomp,color}
\usepackage{booktabs}
\usepackage{multirow}
\usepackage{natbib}
\usepackage{comment}
\usepackage{float}
\usepackage{caption}
\usepackage[section]{placeins}
\graphicspath{{../Figures/}}
\usepackage{verbatim}
\usepackage{psfrag}

\usepackage{hyphenat}

\usepackage{setspace}

\allowdisplaybreaks

\journal{Composites Part B: Engineering}

\usepackage{setspace}

\def\textsim{{\raise.17ex\hbox{$\scriptstyle\sim$}}}

\begin{document}
\begin{frontmatter}
	
\title{\large\textbf{Geometrical and Mechanical Characterisation of Hollow Thermoplastic Microspheres for Syntactic Foam Applications}}

\author[1,2]{Matthew E. Curd}
\ead{matthew.curd@manchester.ac.uk}\corref{cor1}
\cortext[cor1]{Corresponding author}
\author[2]{Neil F.\ Morrison}
\ead{neil.morrison@manchester.ac.uk}
\author[2,3]{Michael J. A. Smith}
\ead{ms2663@cam.ac.uk}
\author[1,2]{Parmesh Gajjar}
\ead{parmesh.gajjar@alumni.manchester.ac.uk}
\author[2]{Zeshan Yousaf}
\ead{zeshan.yousaf@manchester.ac.uk}
\author[2]{William J. Parnell}
\ead{william.parnell@manchester.ac.uk}

\address[1]{\scriptsize{Henry Royce Institute, Department of Materials, University of Manchester, Oxford Road, Manchester, M13 9PL, United Kingdom}}
\address[2]{Department of Mathematics, University of Manchester, Oxford Road, Manchester, M13 9PL, United Kingdom}
\address[3]{DAMTP, Centre for Mathematical Sciences, University of Cambridge, Wilberforce Road, Cambridge, CB3 0WA, United Kingdom}

\begin{abstract}
	\normalsize
	Recently, hollow thermoplastic microspheres, such as Expancel made by Nouryon, have emerged as an innovative filler material for use in polymer-matrix composites. The resulting all-polymer \textit{syntactic foam} takes on excellent damage tolerance properties, strong recoverability under large strains, and favourable energy dissipation characteristics. Despite finding increasing usage in various industries and applications, including in coatings, films, sealants, packaging, composites for microfluidics, medical ultrasonics and cementious composites, there is a near-complete absence of statistical geometrical information for Expancel microspheres. Further, their mechanical properties have not yet been reported. In this work we characterise the geometrical quantities of two classes of Expancel thermoplastic microspheres using X-ray computed tomography, focused ion beam and electron microscopy. We also observe the spatial distribution of microspheres within a polyurethane-matrix syntactic foam. We show that the \textit{volume-weighted} polydisperse shell diameter in both classes of microsphere follows a normal distribution. Interestingly, polydispersity of the shell wall thickness is not observed and in particular the shell thickness is not correlated to the shell diameter. We employ the measured geometrical information in analytical micromechanical techniques in the small strain regime to determine, for the first time, estimates of the Young's modulus and Poisson's ratio of the microsphere shell material. Our results contribute to potential future improvements in the design and fabrication of syntactic foams that employ thermoplastic microspheres. Given the breadth of fields which utilise thermoplastic microspheres, we anticipate that our results, together with the methods used, will be of use in a much broader context in future materials research.	
\end{abstract}


\end{frontmatter}

\section{Introduction}

Syntactic foams are lightweight synthetic composites consisting of a metal, ceramic, or polymer matrix and microsphere inclusions \cite{Orbulov2009,Gupta2004,Luong2014,Dando2018, yousaf2020compression}. The mechanical properties of syntactic foams can be  tailored by adjusting the matrix material and/or the type and volume fraction of the microspheres \cite{Gupta2004,Luong2014,Dando2018}. For these reasons, syntactic foams are suitable for applications within numerous industries including the aerospace, automotive, marine, and sports equipment sectors \cite{Luong2014,Dando2018, gupta2014applications, john2014syntactic,li2008self,Gladysz2015d}. 

In general, syntactic foams exhibit complex loading/unloading behaviour, which becomes more pronounced at higher microsphere volume fractions \cite{dando2018production,Luong2014,Ariff2010, paget2020syntactic}. Under uniaxial compression, the stress-strain response of a typical syntactic foam, with a large (40-50\%) volume fraction of \textit{glass} microspheres, consists of three regimes. At low strains the deformation is entirely linearly elastic, with the microspheres providing a strong stiffening mechanism. At medium strains, the macroscopic response exhibits a plateau region, due to the crushing failure of individual microspheres. At high strains, the material behaves linearly once again due to densification of the microspheres, i.e., the filling of the microsphere cavities with debris. For this reason, glass microsphere foams are non-recoverable over a large strain regime and are therefore inappropriate in applications involving large repetitive strains. Hollow \textit{plastic} microspheres appear to be more suitable under such loadings, as the shells tend to flatten and buckle under load, without fracture~\cite{de2013predicting, yousaf2020compression}. An all-polymer syntactic foam is therefore identified as possessing significantly higher ``recoverability'' than glass-based foams. Despite this attractive property, plastic microsphere foams have received significantly less attention in the research literature  \cite{everett1998microstructure,mae2008effectsexpancelpmma,smith2018analyticaldata,mae2008effects} compared to glass microsphere foams~\cite{zeltmann2017mechanical,dando2018characterization,gupta2006characterization,pinisetty2015hollow,gupta2010comparison,corigliano2000experimental,dando2018production} although uses of thermoplastic microsphere composites have ranged from underwater applications  \cite{martin2007low} to maxillofacial prosthetics  \cite{,liu2015characterization}. Furthermore, recently it has been suggested that a mixture of glass \textit{and} plastic microsphere fillers gives rise to hybrid materials with unusually high filler fractions and associated useful material properties~\cite{dando2019characterization, dando2020nano}.

It is well understood that, in a similar fashion to many reinforced materials, syntactic foams of a given microsphere volume fraction can exhibit differing mechanical properties according to the size, shape and distribution of the filler, i.e.\ the properties of the embedded microspheres \cite{bardella2001elastic, yu2013effects, nian2014effects,d1999analysis, bardella2018failure}. In order to save time, effort, and money in the development of new syntactic foams, it is beneficial therefore to have methods that can \textit{predict} the mechanical response of syntactic foams, given information regarding their constituent parts and microstructure. An important focus in modelling the mechanical behaviour of syntactic foams has been the linear elastic regime; for which both analytical and computational models have been proposed
\cite{bardella2001elastic, Porfiri2009, yu2013effects, nian2014effects, cho2017finite, bardella2018failure}. Computational models are predominantly finite element models of representative volume elements (RVEs) drawn from idealised or imaged microstructures. For example, polydispersity of glass microballoons was incorporated in models of glass-epoxy syntactic foams in \cite{bardella2018failure}. A significant bottleneck which hampers the success of such models however is the availability and quality of the microstructural data used as inputs. 

This study focusses on polyurethane matrix syntactic foams where the filler is \textit{hollow plastic (Expancel) microspheres}. A critical parameter required as an input to associated theoretical and computational models is the \textit{distribution} of the diameter, $a$, of the microspheres. For the Expancel microspheres studied here, an estimate of the median diameter is specified by the manufacturer~\cite{920_diam_shell} but precise information and distributions have not been reported in the literature. Similarly, it is also hitherto unknown whether a homogeneous spatial distribution of microspheres is achieved within the final foams, or if there is residual clustering from the manufacturing process. Clustering of microspheres may lead to anisotropy in the mechanical properties of the final foam and also the potential for lower damage tolerance \cite{yu2013effects}. Previously, microsphere diameter distributions have been obtained through stereography of 2D optical micrographs \cite{Orbulov2009,Carlisle2006}. More recently, X-ray CT has been used to image thermoplastic microspheres, with 2D analysis of the digital CT slices revealing the diameter distribution \cite{Dando2018}.

In addition to microsphere diameter, the microsphere shell thickness affects the mechanical properties of syntactic foams, even when the volume fraction is held constant \cite{Porfiri2009,Ariff2010,wouterson2005specific, bardella2018failure}. The shell thickness, $h$, is therefore another key parameter for any syntactic foam design/modelling scheme. The shells of the Expancel microspheres studied here are expected to be of the order of hundreds of nanometres in thickness and therefore unresolvable using conventional micro-scale X-ray CT. As with the diameters, statistical information regarding shell thickness distributions has not previously been reported in the literature \cite{920_diam_shell}. Previous modelling works have inferred the shell thicknesses of microspheres by considering the sizes and densities of the microspheres~\cite{Gupta2004,Gupta2013,wouterson2005specific}. This approach assumes that the shell thickness of an individual microsphere is a function of its diameter and further that microspheres of a given diameter have the same shell thicknesses. It has not previously been checked as to whether this assumption is valid for the Expancel microspheres studied here, but it was shown to be incorrect for the thick (micrometre-scale) carbon microspheres in \cite{Carlisle2006} where (see Fig.\ 9 of \cite{Carlisle2006}) there appeared to be no correlation between diameter and shell thickness. It is likewise unknown whether the wall thickness of each individual microsphere is uniform \cite{Gupta2004,Gupta2013}.

It should be noted that in a more general setting, the cross-linking and expansion properties of thermoplastic microspheres have been studied and characterised, see e.g.\ \cite{kawaguchi2004synthesis}. Such studies have been performed given the use of the material in a huge range of other applications as a lightweight filler and blowing agent. In these contexts, use of microspheres can reduce cost of manufacture, reduce weight, improve material texture and they can also yield a number of improved material properties for use in coatings \cite{sandin2017reflective, tomalino1997heat}, pressure-sensitive films \cite{chen2020bio}, sealants~\cite{li2015preparation}, packaging~\cite{chan2012novel}, composites for microfluidic applications \cite{samel2007thermally}, medical ultrasonics applications~\cite{zeqiri2000new}, cement binders~\cite{aglan2009strength}, and cementious composites \cite{brooks2018comparative}. 

In the present work, geometrical properties of Expancel thermoplastic microspheres are characterised using X-ray computed tomography (CT), radiography, focussed ion beam (FIB) and scanning electron microscopy (SEM). The spatial arrangement and diameter distribution of in-foam microspheres is measured through micro-scale X-ray CT with automated 3D analyses, whilst the shell thicknesses are measured using nano-scale X-ray radiography. The shell measurements are then validated using FIB-SEM techniques, analogously to previous studies of solid plastic~\cite{Barlow2014} and hollow glass~\cite{Mee2008} microspheres. Finally, we employ the experimentally deduced geometric parameters in polydisperse micromechanical models and fit to experimental data on uniaxial compression in order to infer a range of values for the Young's modulus and Poisson's ratio of the Expancel shells. These results represent the first characterisation of the geometrical and mechanical properties of Expancel microspheres, providing important baselines for the future study and design of composite foam materials made with polymeric microspheres. Given that these microspheres are employed in a rich variety of fields, as described above, we anticipate that our results, provided here in the context of syntactic foams, together with the methods employed, will be of use in a much broader context in future materials research.

\section{Microsphere Diameters and in-foam Distribution} \label{sec2}

The syntactic foams examined here are made using two different grades of hollow thermoplastic \textit{expanded} microspheres. These are the `920 DE 80 d30' and `551 DE 40 d42' Expancel grades (noting that DE refers to \textit{dry expanded}) \cite{920_diam_shell}; hereafter termed `920' and `551' respectively. These microspheres were distributed within a polyurethane matrix at five different volume fractions (2\%, 10\%, 20\%, 30\% and 40\%) for a total of ten different samples (see \cite{yousaf2020compression} for details regarding the manufacture of the samples). In this section, we characterise the diameter distribution and spatial arrangement of the embedded microspheres using X-ray CT on three of the volume fraction samples, 2\%, 10\% and 40\%, for both the 920 and 551 grades, which provides insight into microsphere distribution across a range of volume fractions.

\subsection{Method}

For each grade and volume fraction combination, match-stick samples (approximately 1~-~2~\si{\milli\metre} thickness and of arbitrary height) were scanned using a Zeiss Xradia Versa 520 X-ray CT instrument. Scanning was carried out using a beam energy of 50~\si{\kilo\volt} and current of 80~\si{\micro\ampere}. For each scan, a minimum of 1601~projections were collected using a 4x~optical~lens and 1x1 detector binning, resulting in a voxel size of \textsim1.5~\si{\micro\metre}. To ensure good signal/noise levels in the final reconstruction, the exposure time of each projection (\textsim10~\si{\second}) was set to provide at least 5000~counts; requiring around 5 hours per sample. The exact settings for each scan are provided in Appendix A. Reconstruction was achieved via filtered back\hyp{}projection, using a FDK (Feldkamp Davis Kress) implementation for cone-beam systems \cite{Feldkamp1984}, within XMRecon software. 

Analysis of the CT results was carried out using Avizo 9.7 software. For each dataset, the microspheres were segmented to facilitate automated analysis. Segmentation involves assigning voxels to different labels intended to represent the phases or features within a material (e.g.\ the matrix and microspheres). Segmentation was achieved here using a standard watershed-based method, similar to that presented in \cite{Weis2017}. Label analyses were then performed, giving positional and geometric data for each microsphere, in all six scanned samples. Features with an equivalent diameter below 5~\si{\micro\metre} were removed from the output, as these are finer than the scan resolution (nominally 3~-~4 \texttimes~voxel size). Similarly, segmented features with aspect ratios greater than 2.0 were discounted, as they are likely to be segmentation artefacts rather than microspheres.

\subsection{Results and Discussion}

\begin{figure}[h!]
	\centering
	\includegraphics[width=0.9\textwidth]{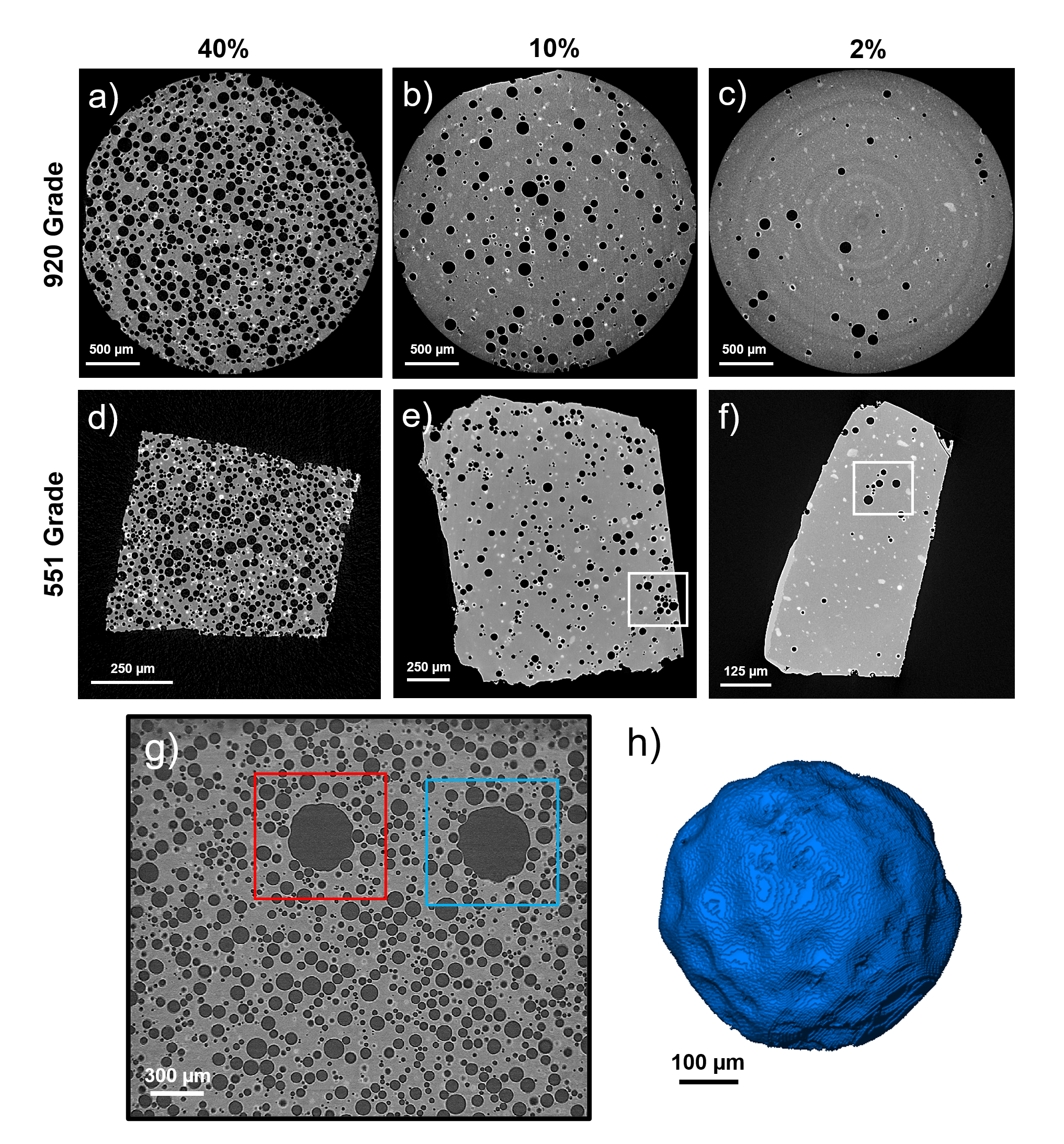}
	\caption{X-ray CT results of the syntactic foam samples: a-f) Examples 2D slices (XY orientation) for each sample, with white boxes highlighting regions where microspheres appear clustered; g) a slice of arbitrary orientation taken from the 40\% volume fraction 920 grade sample, showing two non-spherical features, believed to be air pockets; h) 3D rendering of the feature highlighted in (g) in a blue square.}
	\label{fig::XCTExampleSlices}
\end{figure}

Fig.\ \ref{fig::XCTExampleSlices} shows example 2D slices extracted from the 3D reconstruction of each sample. In these images, the microspheres can be identified as the dark (low density) circular features. 
As anticipated, we observe a distribution in microsphere diameters across all samples, with 920 grade microspheres being generally larger than those for 551 grade. The interior of each microsphere is given by a darker shade, representing the lower density of the encased gas. Surrounding each microsphere there is a brighter shaded ring that arises from in-line phase contrast due to differences in X-ray diffraction from the different materials \cite{Mayo2012}. Inhomogeneity in the spatial distribution of microspheres is apparent in the lower volume fraction samples, where there is a tendency for microspheres to arrange into clusters as highlighted in Fig.\ \ref{fig::XCTExampleSlices}(e,f). The presence of irregular patches in the polyurethane matrix is noted and we conject that these are small inhomogeneities arising from the polyurethane mixing process that will not affect the mechanical behaviour of the samples significantly. It was noted that the 40\% volume fraction 920 grade sample contained several non-spherical features, believed to be air bubbles, as shown in Fig.~\ref{fig::XCTExampleSlices}(g,h). Such features were not found in any of the other samples. It is not known to what extent these bubbles may affect the mechanical properties of the foam. Their presence will not affect the characterisation of the microsphere diameter distribution however, as they were removed from the labels prior to analysis.

\begin{figure}[h!]
	\centering
	\includegraphics[width=0.8\textwidth]{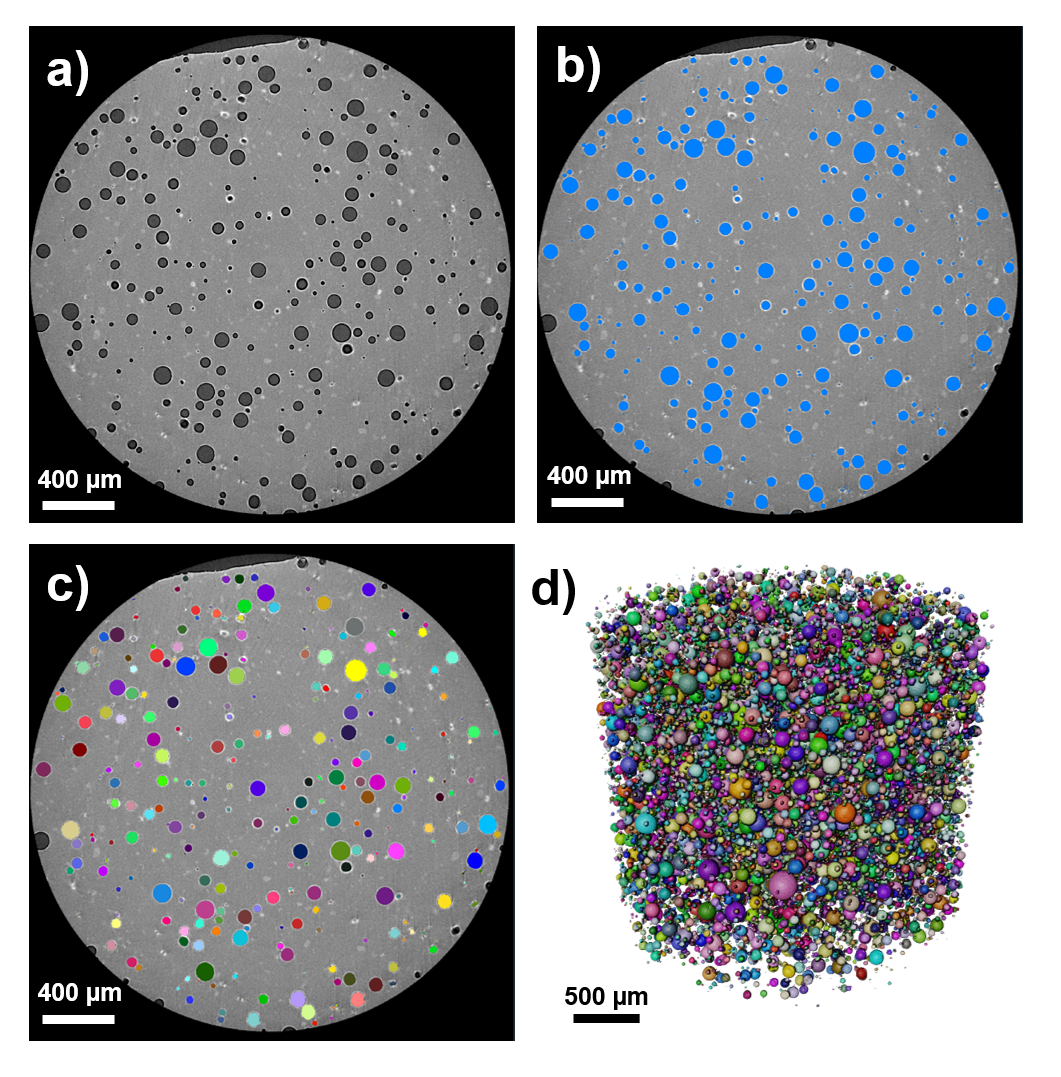}
	\caption{{Depiction of the segmentation and label analysis process: a) example 2D slice of the 10\% 920 grade foam; b) same slice with dark features (microspheres) assigned to a label field via an automatic thresholding process; c) separation of the label field into discrete objects, allowing each microsphere to be independently analysed and/or visualised; d) 3D rendering of the microspheres present in each slice of the CT dataset (as seen from an arbitrary perspective).}}
	\label{fig::3DCT}
\end{figure}

For each scan, the microspheres were segmented to allow automated label analysis, as described in Section 2.1 and depicted in Fig.\ \ref{fig::3DCT}. This was done to give diameter statistics for the microsphere populations within each sample. In each of the 40\% samples we scanned, over 50,000 microspheres were identified and analysed. For the 40\% volume fraction samples we calculate the mean microsphere diameters to be 21.378 \si{\micro\metre} and 36.767 \si{\micro\metre} for the 551 and 920 Expancel grades, respectively.  Averages for the 2\% and 10\% volume fraction foams were also calculated and these were within 5\% of the averages of the 40\% volume fraction foam, for both microsphere grades. As the positions of the match-stick samples were chosen arbritrarily, this suggests little segregation of microspheres according to their size within the original foam moulds.

We present all of our diameter averages in Table \ref{tab:diameters}. To allow comparison with manufacturer specifications, the diameter statistics from the 40\% samples shown in Fig.\ \ref{fig::DiameterStatistics} are presented as \textit{volume-weighted} distributions (see Appendix B for further details). The volume-weighted median value of each grade, computed as 37.744~\si{\micro\metre} and 83.804~\si{\micro\metre} for the 551 and 920 grades, respectively, are both within the ranges provided by the manufacturer, i.e.\ 30~-~50 \si{\micro\metre} for the 551 and 55~-~85 \si{\micro\metre} for 920 \cite{920_diam_shell}. As expected, we note that the mean of the normal distribution of the fit is also slightly different to any of the averages determined on the complete data.

\begin{figure}[h!]
	\centering
	\includegraphics[width=\textwidth]{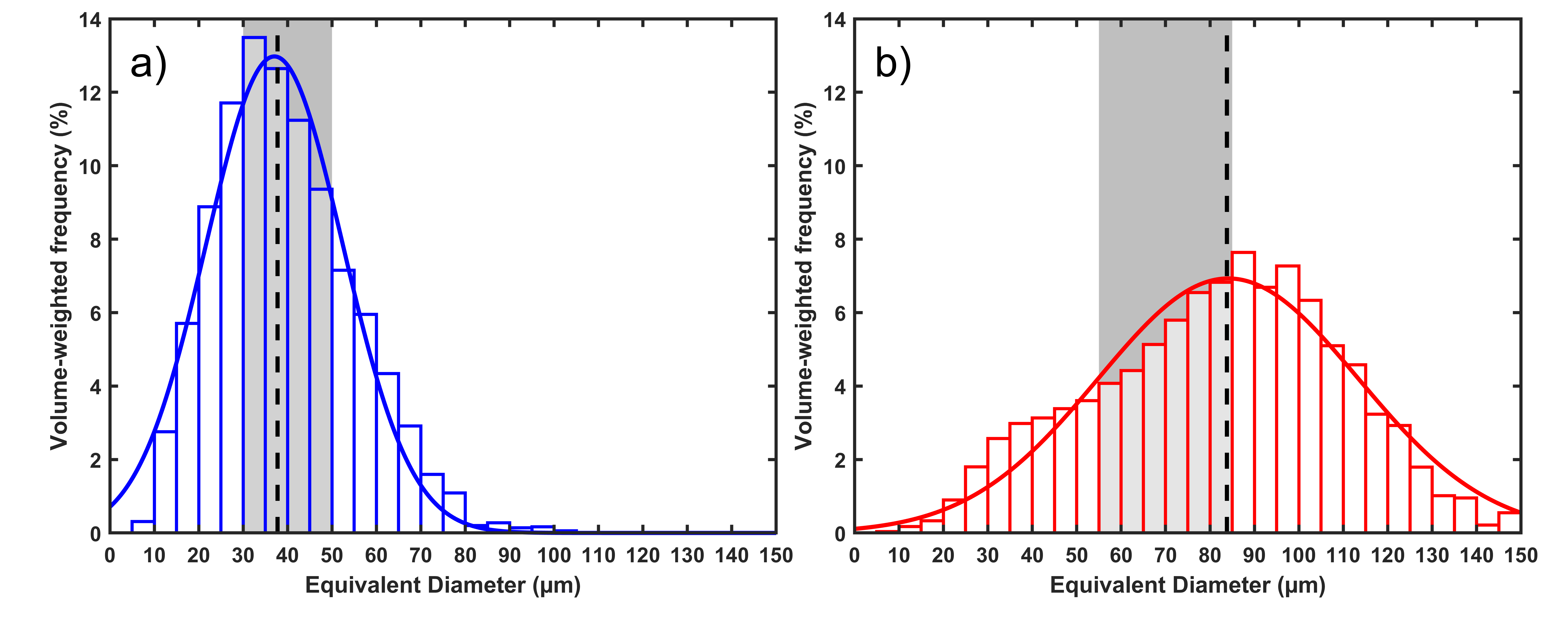}
	\caption{Volume-weighted diameter distributions for the 551 (a) and 920 (b) grade Expancel microspheres within the two 40\% samples. The shaded regions represent the ranges specified by the manufacturer \cite{920_diam_shell} for the volume-weighted median diameter of each grade whereas the dashed lines are our calculations of these values from the segmented CT data, which are 37.744 \si{\micro\metre} and 83.804 \si{\micro\metre} respectively. The volume-weighted diameter, $a$, is distributed as $a$ $\sim$ N($\mu$,$\sigma^{2}$), where N denotes a normal (Gaussian) distribution, with the fitting parameters:  $\mu$(920)=84.1,~$\sigma$(920)=29.3; $\mu$(551)=37.0,~$\sigma$(551)=15.4, where $\mu$ is the mean and $\sigma$ is the standard deviation of the fit.}
	\label{fig::DiameterStatistics}
\end{figure}

\medskip

\begin{table*}[h!]
	\centering
	\begin{tabular}{ccc|ccc}
		\toprule
		\multirow{2}{*}{\textbf{Grade}} & \multicolumn{2}{c}{\textbf{Number-Weighted}} & \multicolumn{3}{c}{\textbf{Volume-Weighted}} \\
		& Mean (CT) & Median (CT) & Mean (CT) & Median (CT) & Median (Manuf.) \\
		\midrule
		
		551 & 21.378 & 18.588 & 39.544 & 37.744 & 30 - 50  \\ 
		
		920 & 36.767 & 30.305 & 81.563 & 83.804 & 55 - 85 \\
		
		\bottomrule
	\end{tabular}
	\caption{Various measures of the average diameter of the 551 and 920 grade microspheres. These values were computed from the segmented X-ray CT data (of the 40\% volume fraction samples) or taken from the manufacturer's specifications \cite{920_diam_shell}, as denoted in parentheses. All units are in \si{\micro\metre}.}
	\label{tab:diameters}
\end{table*}

\section{Microsphere shell thickness} \label{shellthickness}

Nano-scale X-ray radiography was used to measure the shell thicknesses of multiple 920 grade microspheres, of various diameters. To validate these results, FIB sectioning of selected microspheres was performed, with the shell thicknesses recorded through scanning electron microscopy~(SEM).
 
\subsection{Method}

\begin{figure*}[h!]
 	\centering
 	\includegraphics[width=0.9\textwidth]{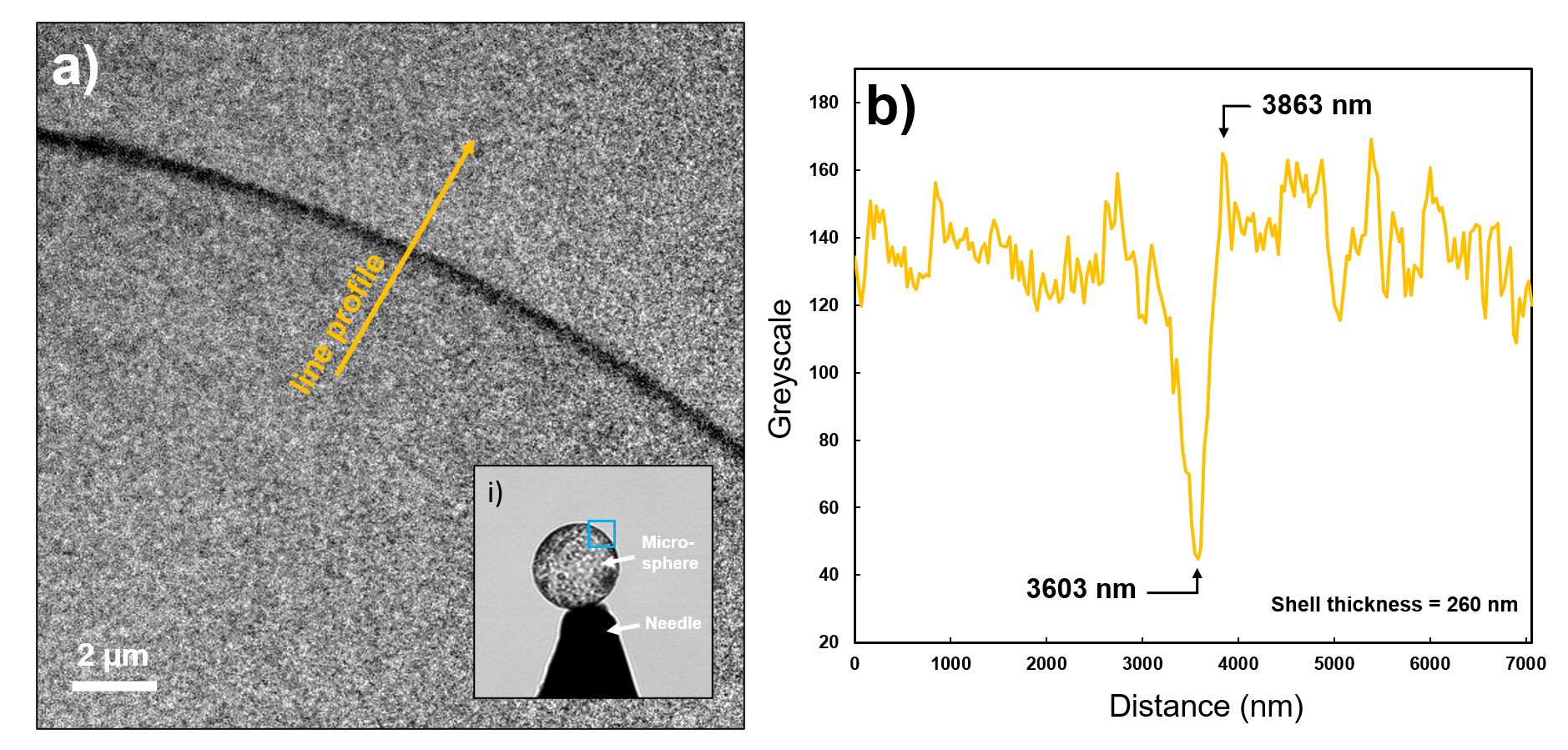}
 	\caption{{Depiction of the shell thickness measurement method: a) X-ray Radiograph showing part of the shell of a hollow plastic microsphere; i) optical image showing the needle\hyp{}mounted sphere and the location where the radiograph was taken; b) 10 pixel wide line profile of the arrow plotted on (a), the shell thickness is taken as the difference between the two highlighted points.}}
 	\label{fig::RadiographyShellThicknessMethod}
 \end{figure*}

 Individual microspheres were picked up electrostatically and then mounted onto the tips of needles with glue. These microspheres were then imaged using a Zeiss Xradia 810 Ultra nano-CT instrument, operating in  high-resolution (16~\si{\nano\meter} pixel size) absorption contrast mode. This instrument uses a parallel beam with a fixed beam energy of 5.4~\si{\kilo\volt} and power of 0.9~\si{\kilo\watt}. For each microsphere, five radiographs (exposure~time $\geq$ 200 s ; 1\texttimes1 binning) were collected and averaged. Fig.\ \ref{fig::RadiographyShellThicknessMethod} depicts the procedure used to infer the shell thicknesses from the radiographs. Line profiles were plotted across the microsphere shell walls and the thicknesses were taken as the difference in position between the point of lowest greyscale and the peak associated with outer shell wall. This method was first tested in MATLAB simulations of X-ray attenutation through an idealised hollow microsphere and was found to be a reliable way to recover a known shell thickness. For each microsphere, the results from up to five unique line profiles were averaged to give a single shell thickness value. The microsphere diameters (generally larger than the radiograph field of view) were recorded using an optical microscope on the same instrument.
 
\sloppy To validate the radiography results, FIB sectioning of five of the same needle\hyp{}mounted spheres was performed using a FEI NovaLab 660 FIB\hyp{}SEM instrument comprising a Ga\textsuperscript{+} ion beam column coupled with a scanning electron microscope (SEM). Prior to sectioning, the needle\hyp{}mounted microspheres were attached to a SEM stub and coated with 15~\si{\nano\metre} of Ag\hyp{}Pd, using a Quorum Q150T sputter coater, to ensure good conductivity. Sectioning of each sphere was done using an ion beam energy of 30~\si{\kilo\electronvolt} and a current of 21~\si{\nano\ampere}, using the ‘Si\hyp{}CCS’ (silicon cleaning cross\hyp{}section) routine. In each case, the final cut line was positioned colinearly to the centre of the microsphere. Regardless of the actual microsphere diameter, the FIB was set to mill to an accumulated depth of 10~\si{\micro\metre} (calibrated against Si.) This was sufficient to fully section microspheres up to 100~\si{\micro\meter} in diameter due to the faster milling of the microsphere material, compared to Si. The cross\hyp{}section of each microsphere was then inclined normal to the electron beam and imaged using the secondary electron (SE) imaging mode, at a beam energy of 5~\si{\kilo\electronvolt} and current of 0.43~\si{\nano\ampere}. 

\subsection{Results and Discussion}

\begin{figure}[h!]
	\centering
	\includegraphics[width=0.75\columnwidth]{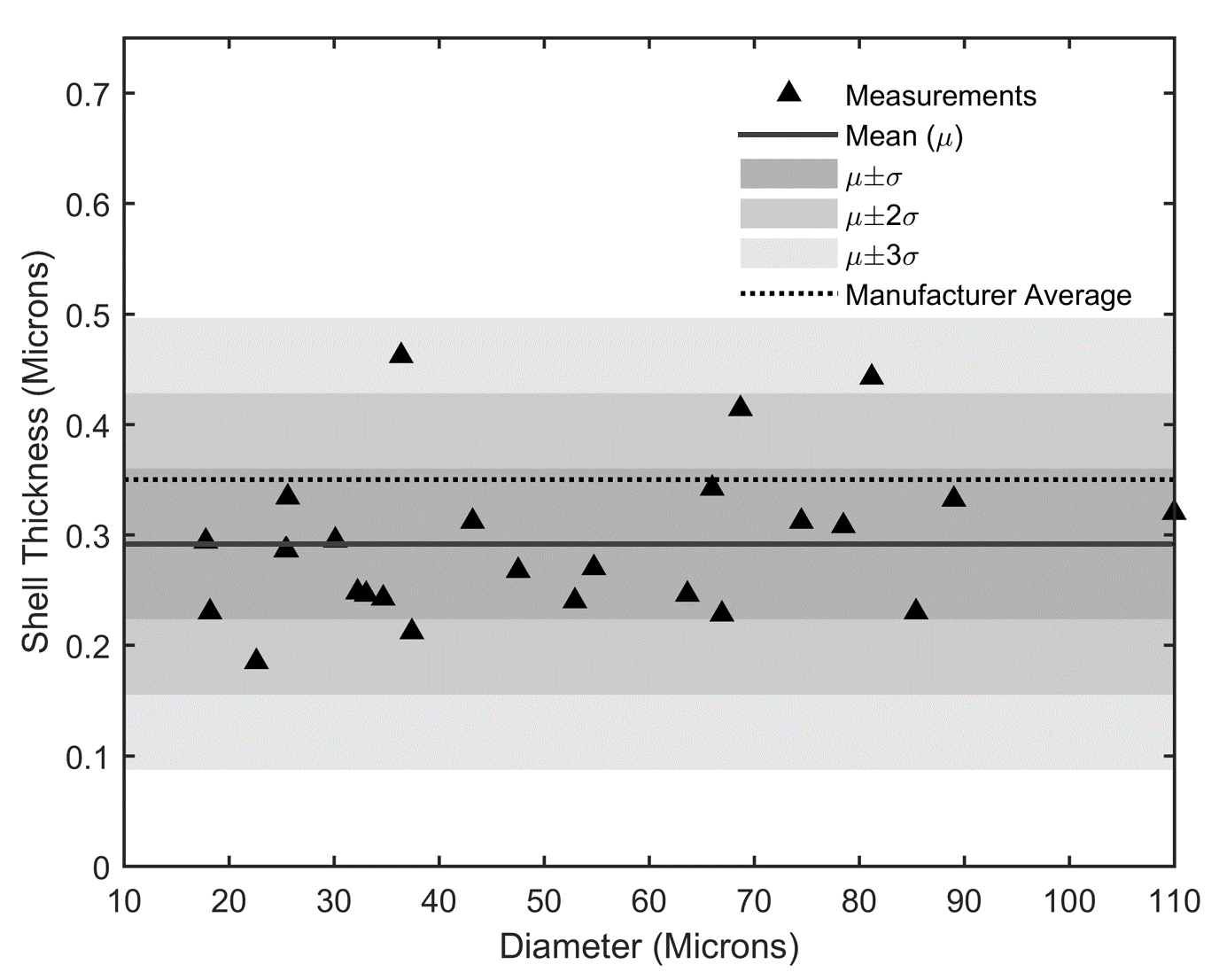}
	\caption{{Summary of the 920 grade shell thickness measurements, from the radiography investigations. The solid line denotes the mean shell thickness from our measurements, whereas the dotted line is an estimate of the mean from the manufacturer \cite{920_diam_shell,Alberts2019}. The shaded regions represent the confidence intervals, derived from the variance of our measurements.}}
	\label{fig::RadiographyShellThicknessResults}
\end{figure}

Radiography was conducted on 25 microspheres (920 grade) and the shell thicknesses were inferred from the radiographs. These results are summarised in Fig.\ \ref{fig::RadiographyShellThicknessResults}. The mean shell thickness (290 nm) and three confidence intervals, derived from the standard deviation of the line profile measurements, are also plotted.  It is clear from \cite{920_diam_shell} that the manufacturer's expectation is that larger microspheres have thinner shells. This has likewise been assumed elsewhere for other types of microsphere in density calculations aiming to infer shell thicknessess, as described in the introduction \cite{Gupta2004,Gupta2013,wouterson2005specific}. Interestingly, we find no correlation between microsphere diameter and shell wall thickness for Expancel microspheres, mirroring results for the thick (micrometre-scale) walled carbon microspheres in \cite{Carlisle2006}. To our knowledge, this is the first time that microspheres with shell thicknesses in the sub-micron range have been measured directly. 

\begin{figure}[h!]
	\centering
	\includegraphics[width=0.85\columnwidth]{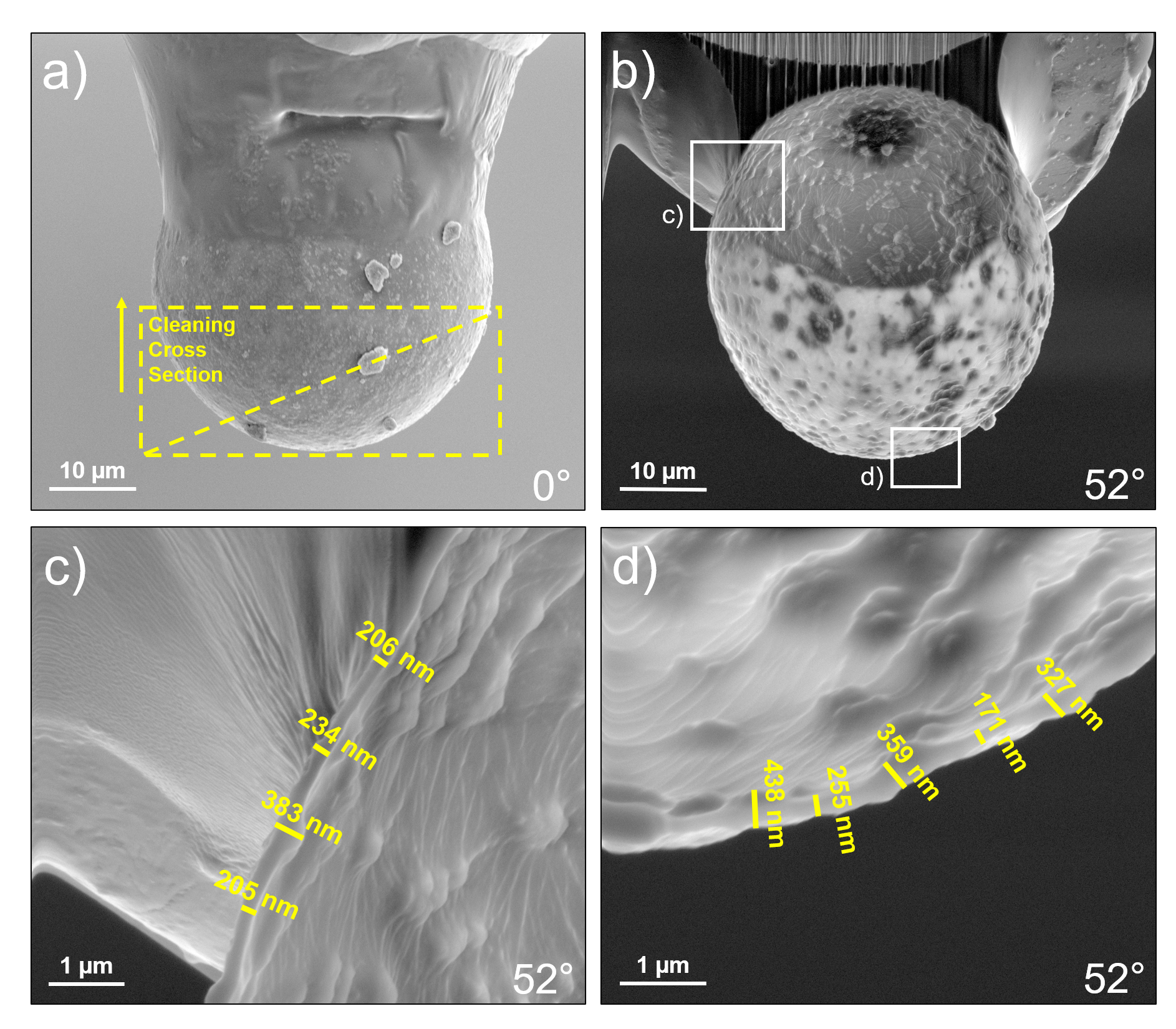}
	\caption{Results of the FIB sectioning investigation: a) secondary electron (SE) image of a Ag-Pd coated microsphere prior to sectioning; b) SE image of the sectioned microsphere, at 52$^\circ$ stage tilt (normal to ion beam); c,d) higher magnification images of the highlighted area, showing the shell thickness measurements; note the variation in shell thicknesses between and within each area.}
	\label{fig::FIBSectioningImages}
\end{figure}

To confirm the shell thickness measurements from the radiography investigations, FIB sectioning of five needle\hyp{}mounted microspheres was conducted. Fig.\ \ref{fig::FIBSectioningImages} shows an example of the SEM images that were collected and the procedure to measure the shell thickness. The SE images collected reveal significant variation in the shell thickness. This  can be as large as 256\%, between points separated by only a few microns, as shown in Fig.\ \ref{fig::FIBSectioningImages}(c,d). The reason for such large local variations is unknown but may be related to variations inherent in the original, unexpanded microspheres and/or the expansion process itself. Table \ref{tab::fib-sem_table} summarises the shell measurements. For each microsphere, the CT and FIB-SEM measurements of the shell thickness are consistent, when considering the uncertainty of each technique.

\begin{table*}
	\centering
	\begin{tabular}{cccccc}
		\toprule
		\multirow{2}{*}{\#} & \textbf{{Diameter}} & \multicolumn{2}{c}{\textbf{FIB\hyp{}SEM}} & \multicolumn{2}{c}{{\textbf{Radiography}}} \\
		& { $a$ (\si{\micro\metre})} & $h$ (\si{\nano\metre}) & $\sigma_h$ (\si{\nano\metre}) & $h$ (\si{\nano\metre}) & $\sigma_h$ (\si{\nano\metre}) \\
		\midrule
		
		1 & 33.5 & 249 & 71 & 212 & 41 \\ 
		
		2 & 40.0 & 261 & 84 & 312 & 26 \\
		
		3 & 45.8 & 369 & 79 & 268 & 39 \\
		
		4 & 69.9 & 245 & 59 & 292 & 61 \\
		
		5 & 79.1 & 230 & 78 & 230 & 25 \\
		\bottomrule
	\end{tabular}
	\caption{Comparison of the FIB\hyp{}SEM and nano-scale radiography measurements of the microsphere shell thickness, $h$, and the associated standard deviation, $\sigma_h$.}
	\label{tab::fib-sem_table}
\end{table*}

\section{Mechanical properties of the thermoplastic shell}

Given the small scale of the Expancel microspheres it has traditionally been very difficult to measure the mechanical properties of the shell material and therefore, estimates of the associated Young's modulus and Poission's ratio are currently not available. Such estimates would provide an important baseline for the understanding and design of syntactic foams and would be applicable to the many other applications where thermoplastic microspheres are used, as described in the introduction. In \cite{shorter2010axial} a value of 3 GPa was quoted in error as the Young's modulus of an Expancel shell. This value appears to be the modulus of a separate material in the cited investigation \cite{trivett2006investigation}, however.  Here, we aim to indirectly determine the Young's modulus of Expancel microsphere shells by using the geometric statistics acquired in Sections 2 and 3 in combination with previously obtained experimental uniaxial compression data on 920 grade syntactic foams in \cite{yousaf2020compression}.

\subsection{Method}

We employed a modified version of the generalized self-consistent method (GSCM) devised originally by Christensen and Lo \cite{christensen1979solutions, christensen1986erratum}, adapted for the case of hollow spherical inclusions by Herv\'{e} and Pellegrini \cite{herve1995elastic}, and extended to handle multiple types and/or sizes of inclusion by Bardella and Genna \cite{bardella2001elastic}. The RVE is a “composite sphere assemblage” in which each isolated three-phase composite sphere (void, shell, matrix) is treated as being surrounded by an unbounded outer phase known as the equivalent homogeneous medium. The effective shear and bulk moduli are defined as ratios of appropriate stress and strain components, volume averaged over the whole RVE \cite{zaoui1997structural}. The benefits of this model over others have been discussed in detail in \cite{bardella2012critical} and this motivated the choice of model for the present polydisperse study. 

We incorporated data obtained in Section \ref{sec2} on the polydisperse nature of the shell diameters into the method and optimised the fit to the measured effective Young's modulus in \cite{yousaf2020compression} by employing the unknown (model) elastic properties of the Expancel shell as fitting parameters. The distribution of shell diameters was represented in the model as a set of 29 discrete types of composite sphere, weighted by volume, as is required in the model due to averaging over shell diameter distribution. For each type, the aspect ratio of the shell (interior shell diameter divided by the exterior shell diameter) in the model was based on the measurements shown in the histogram in Fig.\ \ref{fig::DiameterStatistics}(b). The outer radius of the composite sphere assemblage was determined by imposing that the RVE volume fraction is equal to the global volume fraction, for each sample considered. To quantify the importance of incorporating a distribution of shell diameters, we also considered another RVE consisting of an assemblage of composite spheres of a single diameter/aspect ratio, based on the volume-weighted mean diameter of the distribution, i.e.\ 81.563~\si{\micro\metre} from Table \ref{tab:diameters}.

Initially we fixed the matrix properties as those that were measured in \cite{yousaf2020compression}, i.e.\ $E = 7.08$~MPa, $\nu = 0.49$, although given the almost-incompressible nature of this material, and the high sensitivity of any prediction based on these values, we later allowed these properties to vary within certain constraints, as explained below. We employed a fixed shell thickness of 290 nm (from Fig.\  \ref{fig::RadiographyShellThicknessResults}). In a separate study we employed the distribution of shell thicknesses as measured in Section \ref{shellthickness} and deduced that predictions are almost completely insensitive to the shell thickness dispersion measured. We therefore concluded that the method is robust to this variation. 

We anticipate the GSCM model to be most accurate at dilute to moderate volume fractions. Therefore, in order to determine the mechanical properties of the shell material, the optimisation was performed on the predicted effective Young's modulus of the syntactic foams with 2\%, 10\%, 20\% and 30\% volume fractions. The objective function to be minimized in a surrogate optimization algorithm was defined as the mean squared error between the measurements of the Young’s modulus and the effective Young’s modulus of the model RVE, the latter calculated either for the polydisperse case or the volume-weighted mean case.  All physical parameters were non-dimensionalized.

\subsection{Results and Discussion}

\begin{figure}[h!]
	\centering
	\includegraphics[width=0.8\columnwidth]{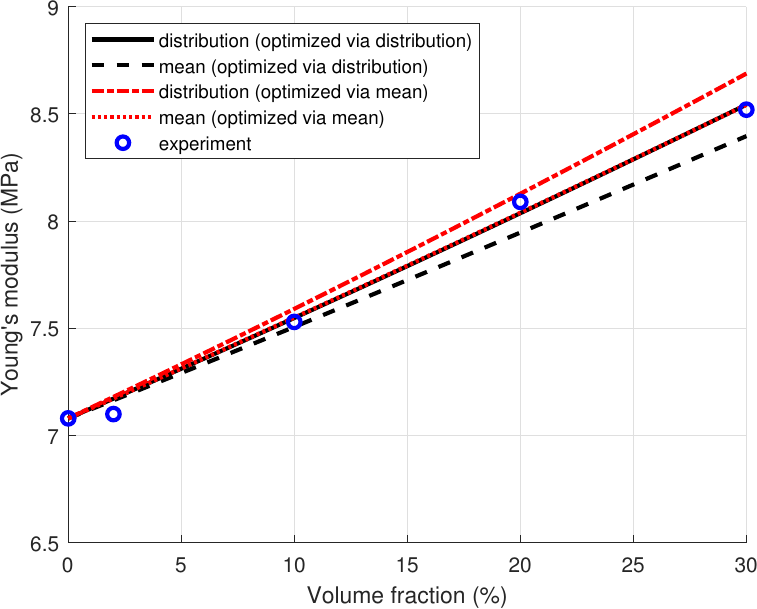}
	\caption{Illustrating the fit to data when matrix properties are fixed as $E_m=7.08$ MPa, $\nu_m=0.49$ and we allow shell properties to vary. Two cases for the fit are employed: one which employs the volume-weighted mean diameter of $81.563$ \si{\micro\metre} and one that uses the full polydispersity of the shells. 
	}
	\label{fig:properties}
\end{figure}

Initially we employed the volume-weighted mean diameter of 81.563 \si{\micro\metre} in order to provide a prediction of the shell properties. Given prior information about the thermoplastics from which the shells are fabricated \cite{arkema2002}, we searched within the following ranges: $1.5<E_s<3$~GPa and $0.2<\nu_s<0.45$.  With reference to Fig.\ \ref{fig:properties}, this identified an optimal fit to data with $E_s\approx~1.78275$~GPa, $\nu_s\approx0.200277$. Using these same bounds on shell properties, we employed the full diameter dispersion data in the GSCM, which yielded predictions of  $E_s\approx 1.67518$ GPa, $\nu_s\approx 0.201136$. Next, reflecting the fact that the matrix is almost incompressible, and therefore even small errors in the measurement of its Poisson's ratio can lead to large errors in the fit, we also allowed the matrix properties to vary in the range $7.075<E_m<7.085$ MPa, i.e.\ to within the rounding error of the experimental measurement, and
$0.485<\nu_m<0.499$ allowing a slightly wider range for this given the sensitivity of properties in this parameter range, and we performed the same fit to data. This yielded values as summarised in Table \ref{tab:properties} and we are therefore able to specify the following ranges of values for the mechanical properties of the shell material:  $1.675<E_s<1.826$ GPa and $0.2<\nu_s<0.275$. 

\begin{table*}[h!]
	\centering
	\begin{tabular}{cccccc}
	\toprule
Distribution & Properties varied & $E_m$ (GPa) & $\nu_m$ & $E_s$ (GPa) & $\nu_s$ \\
\midrule
Mean 			& $E_s, \nu_s$ & 0.00708 & 0.49 & 1.78275 & 0.200277  \\ 				 
Polydisperse	& $E_s, \nu_s$ & 0.00708 & 0.49 & 1.67518 & 0.201136 \\
Mean			& $E_m,\nu_m,E_s,\nu_s$ & 0.00707503 & 0.494626 & 1.82630 & 0.239699\\
Polydisperse 	& $E_m,\nu_m,E_s,\nu_s$ & 0.0070758 & 0.495913 & 1.74987 & 0.274950\\
\bottomrule	
	\end{tabular}
	\caption{Optimised estimates for the Young's modulus $E_s$ and Poisson's ratio $\nu_s$ of the shell material assuming mean and polydisperse shell distributions with fixed elastic properties for the matrix. Also given are optimised estimates for both varied matrix and shell properties in these two shell diameter distribution settings.}
	\label{tab:properties}
\end{table*}

Fig.\ \ref{fig:properties} illustrates the fit to the data in the case when the matrix properties are fixed. The curves are indistinguishable from those given when matrix properties are permitted to vary and therefore we do not show these curves. In Fig.\ \ref{fig:properties} we indicate curves obtained in order to fit to the data, either via optimisation on the mean shell diameter or the polydisperse case. Once this fit has been achieved, we also plot the resulting prediction of effective properties based on the polydisperse distribution or mean diameter respectively given this initial fit to data. We note that the resulting predictions of the shell mechanical properties are relatively close when using the volume-weighted mean diameter and polydisperse diameter distribution. Given that employing the mean diameter in the model is considerably quicker than when employing the polydisperse distribution (of the order of 40 times quicker for the 29 bin case employed here), one may conclude therefore that employing the mean is significantly more efficient. There are two caveats to this however. The first is that this calculation was only required once, i.e. to predict the shell mechanical properties, so it would seem sensible to include as much information as possible in order to be as accurate as we can be for such a calculation. The second is that for a different polydisperse distribution, the prediction arising from the volume-weighted mean may not capture the correct behaviour. This would always need to be checked against the polydisperse result, rather than assuming such a result a priori.

\section{Discussion}

The principal reason for the development of composite materials is to unlock material properties that are far beyond what is possible when employing standard, homogeneous media. However these properties are limited by the constituent parts comprising the composite in question, together with the associated materials fabrication mechanisms. Improvements in composite manufacturing techniques as well as careful characterisation of the building blocks of the materials in question are thus both critical in order to improve composite design. The work described here to characterise the geometrical and mechanical properties of Expancel microspheres therefore contributes significantly to the goal of the improved design of syntactic foams with thermoplastic microsphere fillers. Furthermore, as described in the introduction, given the extensive use of these microspheres in a broad range of applications, their improved characterisation serves a much broader goal in terms of materials design far beyond syntactic foams. 

The success of the characterisation methods employed here highlights their potential for use in experimentally-informed, physics-based materials models to accurately predict the behaviour of existing syntactic foams and other materials systems. They will also assist in the digital design of new foams with bespoke material properties, providing significant enhancements over ad-hoc experimental development of composites design and simultaneously leading to a more sustainable approach to materials design. In recent years, novel computational tools have emerged with the potential to radically change our capacity to optimise the design of composites. These methods are based on supervised machine learning techniques, with the supervised learning based on physics-based modelling of simpler systems. Such methods, developed for now on relatively simple systems, with a reduced parameter space, have nevertheless provided enhanced designs that could not be achieved by brute-force, exhaustive methods \cite{gu2018novo, chen2019machine}. Similar techniques still require significant work to be applicable for improved and optimised design of materials such as syntactic foams, given the distribution of diameters of the microspheres fillers. What can be said however is that even when such improvements are made to these methods, they will be limited by the knowledge of the properties of the constituent parts of the composite materials in question. This illustrates the importance of the characterisation tools described here. Coupled with advances in machine learning techniques, such characterisation tools show exciting promise for the future of composite materials and more generally for materials design.   

\section{Conclusions}

In this work, the geometric and mechanical properties of Expancel thermoplastic microspheres embedded within polyurethane matrix syntactic foams were investigated for the first time, using a combination of X-ray Computed Tomography, nano-scale X-ray radiography and secondary (SE) imaging with focussed ion beam (FIB) sectioning. X-ray CT revealed that the 551 and 920 grades of Expancel microspheres exhibit normal volume-weighted diameter distributions. Volume-weighted median values were computed as 37.744 \si{\micro\metre} and 83.804 \si{\micro\metre}, respectively, both of which lie within the manufacturer specifications, although the range of those stated specifications were relatively wide. The tomography also revealed that microspheres have a tendency to cluster together within the foams, particularly in the dilute (2\% and 10\%) cases and this could potentially lead to local anisotropy in the material properties. Nano-scale radiography was conducted on individual microspheres and revealed a high degree of variation in the shell thicknesses. However, no correlation was found between the shell thickness and microsphere diameter. FIB sectioning of microspheres allowed for direct observations (using SE imaging) of the microsphere shell walls. The images verified our radiography measurements and demonstrated the non-uniformity of the microsphere shells.

Finally, we employed the experimentally deduced geometric parameters in polydisperse micromechanical models and we fitted these models to pre-existing experimental data on uniaxial compression in order to infer a range of values for the Young's modulus and Poisson's ratio of the Expancel shells, noting that these properties have not previously been reported in the literature.  As indicated above, the success of these characterisation methods can be critical when employed in future digital designs of syntactic foams and other innovative materials systems.
 
\section*{Acknowledgements}

The authors are grateful to the Engineering and Physical Sciences Research Council (EPSRC) for funding via grants EP/L018039/1 and EP/S019804/1. We thank Alison Daniel (Thales UK) for sample manufacturing. Beamtime was kindly provided by the Henry Moseley X-ray Imaging Facility (HMXIF), which was established through EPSRC grants EP/F007906/1, EP/I02249X/1 and EP/F028431/1. HMXIF is a part of the Henry Royce Institute for Advanced Materials, established through EPSRC grants EP/R00661X/1, EP/P025498/1 and EP/P025021/1.

\bibliography{references_will_short, references_short}
\bibliographystyle{ieeetr}

\appendix

\section{Detailed X-ray CT settings} \label{settings}

As noted in Section 2.1, some experimental settings varied between X-ray CTscans. This was done to account for differences in sample size, sample positioning (relative to the source/detector) and source brightness variation between different days of scanning, or simply to better take advantage of the equipment's availability (e.g.\ adding more projections). Table \ref{tab:settings} gives details of those settings which varied between scans.

\begin{table*}[h!]
	\centering
	\begin{tabular}{ccccccc}
		\toprule
		Grade & Volume \% & Source-Sample & Sample-Detector & Voxel Size & Exposure & Projections \\
		\midrule
		551 & 2		& 7.03 \si{\milli\metre}	& 10.42 \si{\milli\metre}	& 1.36 \si{\micro\metre}	& 8.0 \si{\second}	& 1601 \\
		551 & 10	& 6.01 \si{\milli\metre}	& 10.00	\si{\milli\metre}	& 1.27 \si{\micro\metre}	& 8.0 \si{\second}	& 1601 \\
		551 & 40	& 6.72 \si{\milli\metre}	& 10.42	\si{\milli\metre}	& 1.32 \si{\micro\metre}	& 8.0 \si{\second}	& 1601 \\
		920 & 2		& 9.08 \si{\milli\metre}	& 13.00	\si{\milli\metre}	& 1.39 \si{\micro\metre}	& 10.0 \si{\second}	& 2401 \\
		920 & 10	& 10.20 \si{\milli\metre}	& 14.16	\si{\milli\metre}	& 1.41 \si{\micro\metre}	& 10.0 \si{\second}	& 3201 \\
		920 & 40	& 9.16 \si{\milli\metre}	& 11.16	\si{\milli\metre}	& 1.52 \si{\micro\metre}	& 8.0 \si{\second}	& 2001 \\
		\bottomrule
		\end{tabular} \\
	\caption{Settings for all six CT scans. See Section 2.1 for settings common to all scans, e.g.\ beam energy, beam power.}
	\label{tab:settings}
\end{table*}

\section{Number-weighted diameter distributions} \label{dists}

Particle size data can be summarised using two different types of distribution: number-weighted distribution and volume-weighted distribution, with each particle-size instrument favouring a particle distribution depending on what is measured. Microscopy methods, such as optical microscopy methods, tend to favour number-weighted distributions, whereas laser-diffraction methods (as used by the Expancel manufacturer) report volume-based distributions. Microscopy methods sometimes offer the capability to transform between a number-weighting and volume-weighting, but this relies on an estimate of the particle volume (usually as a sphere). Due to its 3D nature, X-ray CT provides measurements of the exact volume for each particle regardless of the shape, allowing accurate conversion between number-based and volume-based distributions \cite{Gajjar2020}.

Consider a family of $N$ particles, with the $i$th particle having size $d_i$ and volume $V_i$. Then the total volume $V$ is given by

\begin{equation} 	V = \sum_{i=1}^N V_i. \end{equation}

Suppose the distribution is calculated using a series of bins $B_j$ corresponding to particles with diameters in the interval $[s_j,s_{j+1})$. Then the number frequency $f_j^n$ for each bin $B_j$ is given by

\begin{equation} f_j^n = \sum_{i: s_j\leqslant d_i < s_{j+1}} \dfrac{1}{N}, \end{equation}
whereas the volume frequency $f_j^v$ is given by:

\begin{equation} f_j^v = \sum_{i: s_j\leqslant d_i < s_{j+1}} \dfrac{V_i}{V}. \end{equation}

\begin{figure}[h]
	\centering
	\includegraphics[width=\textwidth]{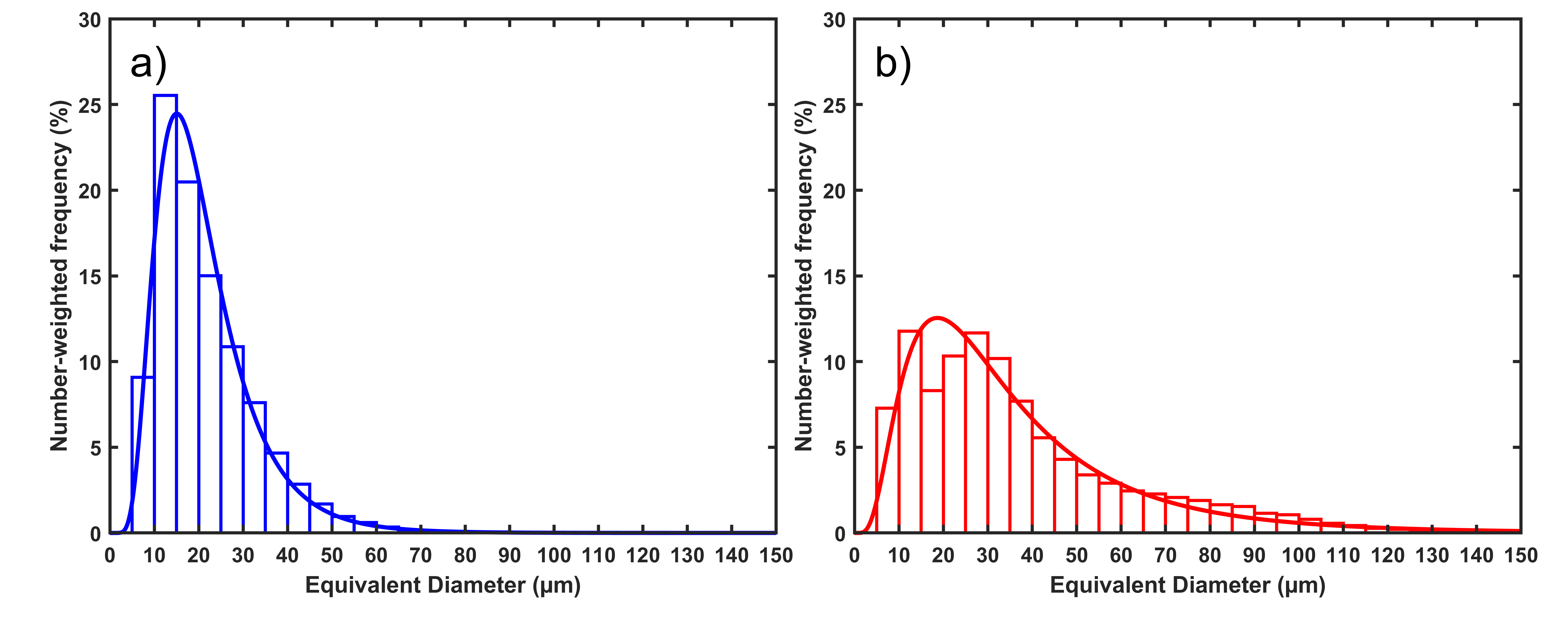}
	\caption{{Number-weighted diameter distributions of Expancel 551 (a) and 920 (b) microspheres. The diameters, $a$, are lognormally distributed as ln($a$) $\sim$ N($\mu$,$\sigma^{2}$), where N denotes a normal (Gaussian) distribution, with the fitting parameters:
	$\mu$(551)=2.944,~$\sigma$(551)=0.482; $\mu$(920)=3.386,~$\sigma$(920)=0.677, where $\mu$ is the mean of the logarithmic diameter values and $\sigma$ is the standard deviation of the logarithmic diameter values.}}
	\label{fig::NumberWeightFinal}
\end{figure}

In a number-weighted distribution, all particles have equal weighting, whereas a volume-weighted distribution favours larger particles since they occupy a greater proportion of the particulate volume. This can be seen by comparing the volume-weighted distribution in Fig.\ \ref{fig::DiameterStatistics} with the number-weighted distribution in Fig.\ \ref{fig::NumberWeightFinal}, where the mean values for both grades are smaller when number-weighted.  The distribution that is required in micromechanical models where averaging takes place over a polydisperse distribution of microspheres naturally turns out to be the volume-weighted distribution and hence this is the focus of the main body of work in the present article.

\end{document}